\pdfoutput=1

\documentclass[10pt,twocolumn,twoside]{pnas-new}

\templatetype{pnasresearcharticle} 

\usepackage{xparse}

\usepackage{appendix}
\usepackage{graphicx}
\graphicspath{ {./} }

\usepackage{dcolumn}
\usepackage{bm}
\usepackage{hyperref}
\usepackage{xcolor} 

\usepackage[noabbrev]{cleveref}

\usepackage{comment}
\usepackage{dsfont}

\newcommand{\ket}[1]{\ensuremath{\left| #1 \right>}}
\newcommand{\bra}[1]{\ensuremath{\left< #1 \right|}}

\usepackage{slashed}

\usepackage{bm}

\usepackage{scalerel}

\usepackage{cancel}
\usepackage{nicefrac}
\usepackage{xfrac}
\usepackage{textcomp}
\usepackage{romannum}
\usepackage{mathtools} 

\usepackage{bbold}

\setboolean{displaywatermark}{false}

\newcommand{\acknowledgements}{\section*{ACKNOWLEDGMENTS}}

\begin{document}


\title{Ab Initio Random Matrix Theory of Molecular Electronic Structure}

\author[a]{Zhen Tao}
\author[b]{Victor Galitski}

\affil[a]{Department of Chemistry, University of Rhode Island, Kingston, Rhode Island, 02881, USA}
\affil[b]{Joint Quantum Institute, Department of Physics, University of Maryland, College Park 20742}


\begin{abstract}
{\bf We use ab initio electronic-structure methods to investigate random-matrix theory (RMT) universality in molecular electronic structure. Using single-reference electronic structure methods, including Hartree-Fock, configuration-interaction singles (CIS), density functional theory, and linear-response time-dependent density-functional theory (LR-TDDFT), we compute single-particle orbital energies and many-electron excitations of several representative molecules (benzene, alanine, 1‑phenylethylamine, methyloxirane, and helicene chains). For generic low-symmetry geometries, the unfolded spectra of these ab initio Hamiltonians exhibit Wigner-Dyson level statistics of the Gaussian orthogonal ensemble (GOE). For extended helicene chains we explicitly restrict to bound valence excitations below the ionization threshold and still observe GOE statistics, indicating that the RMT universality 
is present for physical states of direct relevance to real molecules. We further explore the electric and magnetic field dependence of the  molecular electronic spectra. The variance of electric polarizability (level curvature, $K$) is predicted to be non-analytic in the magnetic field which serves as an infrared cutoff, $\left\langle K^2 \right\rangle \propto \log(1/|B|)$. We observe a transition to the Gaussian unitary ensemble (GUE) by increasing the magnetic fields, although it occurs only at magnetic fields far beyond experimentally accessible scales. Our results indicate that random matrix universality provides a general framework for organizing ab initio predictions of interacting electron spectra in complex systems.} 
\end{abstract}

\maketitle
\par\noindent\textbf{}\theabstract\par\vspace{12pt}

 Electronic structure of molecules and materials is governed by the many-body Schr{\"o}dinger equation, which in principle fixes their structure, spectroscopy, and dynamics. In practice, electron motion in all but the simplest molecules is expected to be chaotic~\cite{Gutzwiller}. Indeed, already three-body  classical  Coulombic problem is non-integrable~\cite{Poincare} and it remains such for more complex $(n>3)$-body problems, which includes the classical limit of just about all molecules of practical interest. Per the Bohigas-Giannoni-Schmit (BGS) conjecture~\cite{BGS_1984} this should lead to quantum chaos and the emergence of the Wigner-Dyson (WD) level statistics of the  molecular electronic spectra~\cite{Mehta,Haake}. On the one hand, this suggests a practical ``complexity barrier'' for accurately predicting individual highly excited electronic levels, because uncertainties in nuclear positions and environment~\cite{DysonBM} render such levels experimentally indistinguishable beyond a certain scale. On the other hand, Random Matrix Theory (RMT)~\cite{Mehta} provides a wealth of universal results that may be useful to computational quantum chemistry applications.
%
%

\begin{figure}[!htbp]
\centering\includegraphics[width=1.0\linewidth]{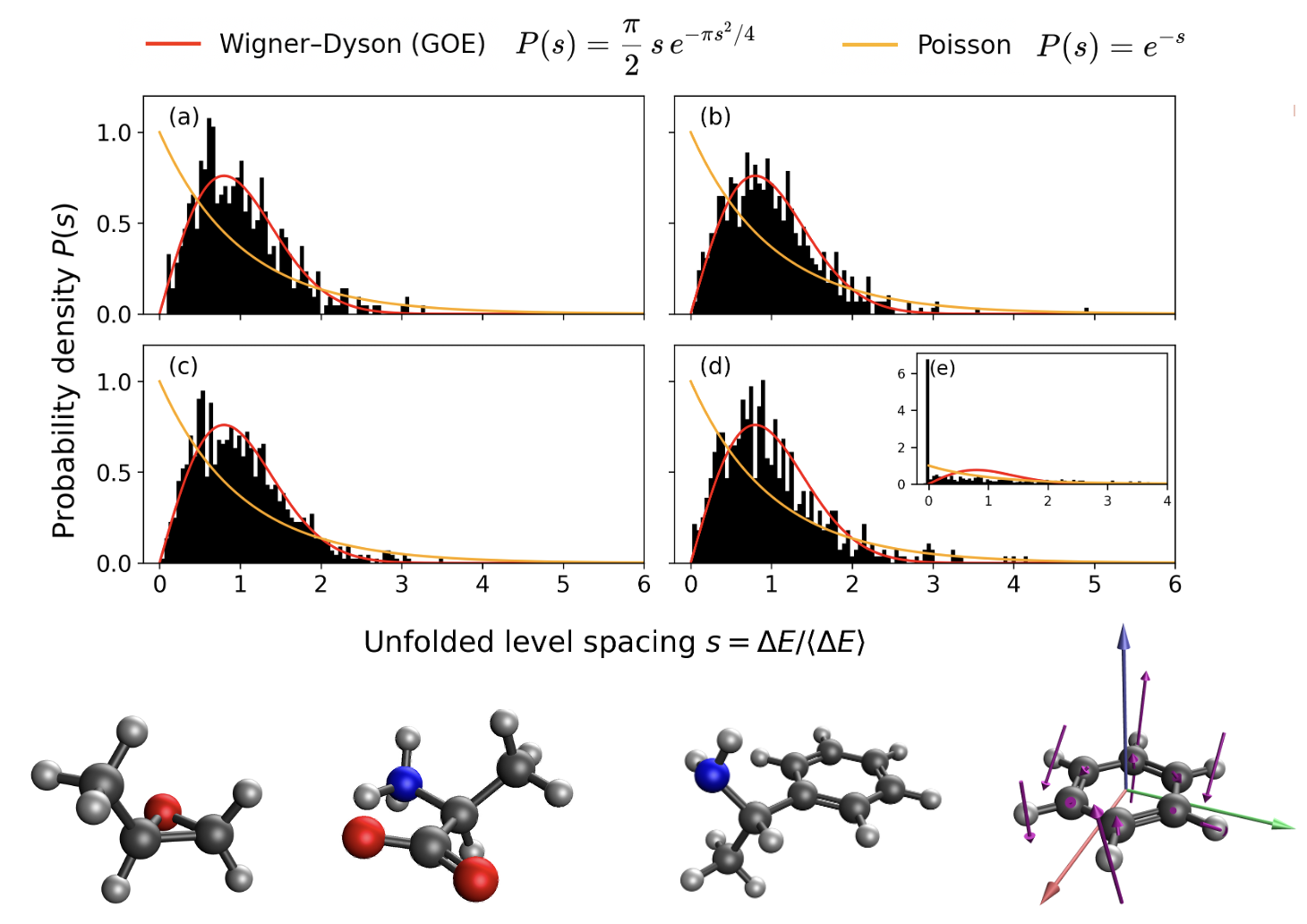}
\caption{Nearest-neighbor level spacing histograms of the unfolded single-particle energies for (a) methyloxirane, (b) alanine, (c) 1‑phenylethylamine, (d) \(C_{1}\)  benzene, and (e)  \(D_{6h}\)  benzene. The molecular structures are shown from left to right in the same order and share the same Cartesian system (x,y,z) with identical axis orientations. The magenta arrows represent the atomic displacement in benzene, which reduces its group symmetry from  \(D_{6h}\)  to \(C_{1}\).  Carbon atoms are shown in dark gray, hydrogen in light gray, nitrogen in blue, and oxygen in red.   }
\label{fig:sp_GOE}
\end{figure}

RMT is a successful theory that was originally developed to capture universal spectral fluctuations in complex nuclei~\cite{Wigner}. It has since become a useful tool for studies of complex quantum systems including  mesoscopic physics~\cite{Beenakker,AZ,wasserman_investigating_2008}, quantum circuits~\cite{Swingle,Google,GalitskiZoller,SFFexp}, photonic systems~\cite{AA,Xanadu,Hiding}, and strongly-correlated  high-energy~\cite{MSS} and condensed matter physics~\cite{SY}. In molecular systems, RMT has been tested against molecular rovibrational and vibronic spectra, where level statistics exhibit signatures of quantum chaos~\cite{abramson_stimulated_1984,persch_vibronic_1988,polik_eigenstateresolved_1990,zimmermann_confirmation_1988}. RMT has since been used to characterize vibrational couplings and vibronic states involving a few low-lying electronic levels~\cite{leitner_statistical_1996,takatsuka_quantum_2022}. In parallel, molecular reaction rate theories such as Rice-Ramsperger-Kassel-Marcus (RRKM)\cite{RobinsonHolbrook1972} rely on assumptions of ergodicity and statistical energy redistribution within the reactant well. Because vibrational couplings in molecules are typically local and structured, local RMT models have been developed to describe vibrational energy flow and intramolecular vibrational energy redistribution~\cite{wolynes_randomness_1992,leitner_quantum_2015,leitner_molecules_2018}. In contrast, the statistical properties and mixing patterns of electronic states for molecular systems have been less systematically examined within a random-matrix framework, despite the potentially more extended nature of electronic Coulomb and exchange interactions.


This paper uses ab initio wavefunction and density functional
electronic structure calculations to examine the BGS conjecture in quantum chemistry  on the examples of several common molecules. We specifically study electronic spectra of benzene, alanine, methyloxirane, 1-phenylethylamine, and complex helicene chains. We find Wigner-Dyson Gaussian orthogonal ensemble (GOE) level statistics in the absence of external fields for all low symmetry geometries. In the case of higher-symmetry molecules (benzene and helicene chains) we recover GOE statistics by displacing the atoms from their equilibrium positions along molecular dynamic trajectories or through atom substitutions that break point group symmetries that otherwise obscure the universal distribution. In the presence of an ultrastrong magnetic field the spectra of all molecules show Gaussian unitary ensemble (GUE) statistics. We further study the random matrix structure of the spectrum in a varying external electric field
and investigate the statistics of level velocities and curvatures, which can be related directly to experimental observables such as transition moments and polarizabilities correspondingly. The infrared limit of the statistical distribution of molecular polarizabilities is predicted to be universal and non-analytic in magnetic field.


\section*{Results and Discussion}

\subsection*{Single-particle quantum chaos} 

We start by investigating single-particle quantum chaos in the electronic structures of a set of small molecules. Within the Born-Oppenheimer (BO) framework,\cite{Born1927,Born1955} the electronic structure for state $I$ is determined by solving the electronic Schr\"odinger equation at each nuclear configuration $\bm R = (\bm R_1, \cdots, \bm R_{N_n})$.
\begin{align}
&\hat H_{\rm el}(\bm r;\bm R)\Psi_I(\bm r;\bm R)
  = E_I(\bm R)\Psi_I(\bm r;\bm R),\quad \text{with} \label{eq:Hel}\\
 &\hat H_{\rm el}(\bm r;\bm R)
  = \sum_{i=1}^{N_e}\frac{\hat{\bm p}_i^{\,2}}{2m_e}
     + \hat V_{\rm ee}(\bm r)
     + \hat V_{\rm en}(\bm r,\bm R)
     + V_{\rm nn}(\bm R) \phantom{............} \notag
\end{align}
 being the non-relativistic electron Hamiltonian, where $\bm r = (\bm r_1, \cdots, \bm r_{N_e})$ denotes the electronic coordinates. $\hat{\bm p}_{i} = -i\hbar \nabla_{\bm r_{i}}$ is the momentum operator for the $i$-th electron and $\hat V_{\rm ee} (\bm r)$, $\hat V_{\rm en} (\bm r, \bm R)$, and $ V_{\rm nn} (\bm R)$ denote the electron-electron, the electron-nuclear, and the nuclear-nuclear Coulomb interactions, respectively.  At the Hartree-Fock (HF) level, the ground state many-electron wavefunction $\Psi_{0}(\bm r;\bm R)$ is approximated by a single Slater determinant of single particle orbitals $\{\psi_i (\bm r)\}$. Applying the Rayleigh-Ritz variational principle~\cite{szabo_modern_1989} yields a set of effective single particle-equations for the orbitals $\psi_i(\bm r)$ and their corresponding energies $\epsilon_i$, which need to be solved self-consistently. 
\begin{align}
    \left[\frac{\hat{\bm p}^2}{2m_e}+ V_{\rm eff}(\bm r) \right] \psi_i(\bm r) = \epsilon_i \psi_i (\bm r),
\end{align}
 $V_{\rm eff}(\bm r)$ is the effective mean-field potential due to electron-nuclear and electron-electron interactions.

We calculated the HF orbital energies for three small chiral molecules (alanine, methyloxirane, and 1-phenylethylamine), a high-symmetry \(D_{6h}\) benzene structure, and a fully symmetry-broken \(C_{1}\) benzene structure generated by perturbing the molecule along a combination of its vibrational modes 1, 3 and 6. In Fig.~\ref{fig:sp_GOE}, we plot histograms of the nearest-neighbor spacings of the unfolded single-particle energies of these molecules (lower panel).
We observe WD GOE statistics for the three chiral molecules in Fig.~\ref{fig:sp_GOE}~(a)--(c) and for the symmetry-broken benzene in Fig.~\ref{fig:sp_GOE}~(d), but not for the highly symmetric benzene in Fig.~\ref{fig:sp_GOE}~(e). This is because the Hamiltonian of the high-symmetry benzene structure decomposes into decoupled 
symmetry 
blocks, each of which is expected to exhibit WD symmetry, whose overlapping spectra obscure level repulsion. Displacing the atoms mixes them up and restores a single WD block.



\begin{figure}[t]
\centering\includegraphics[width=1\linewidth]{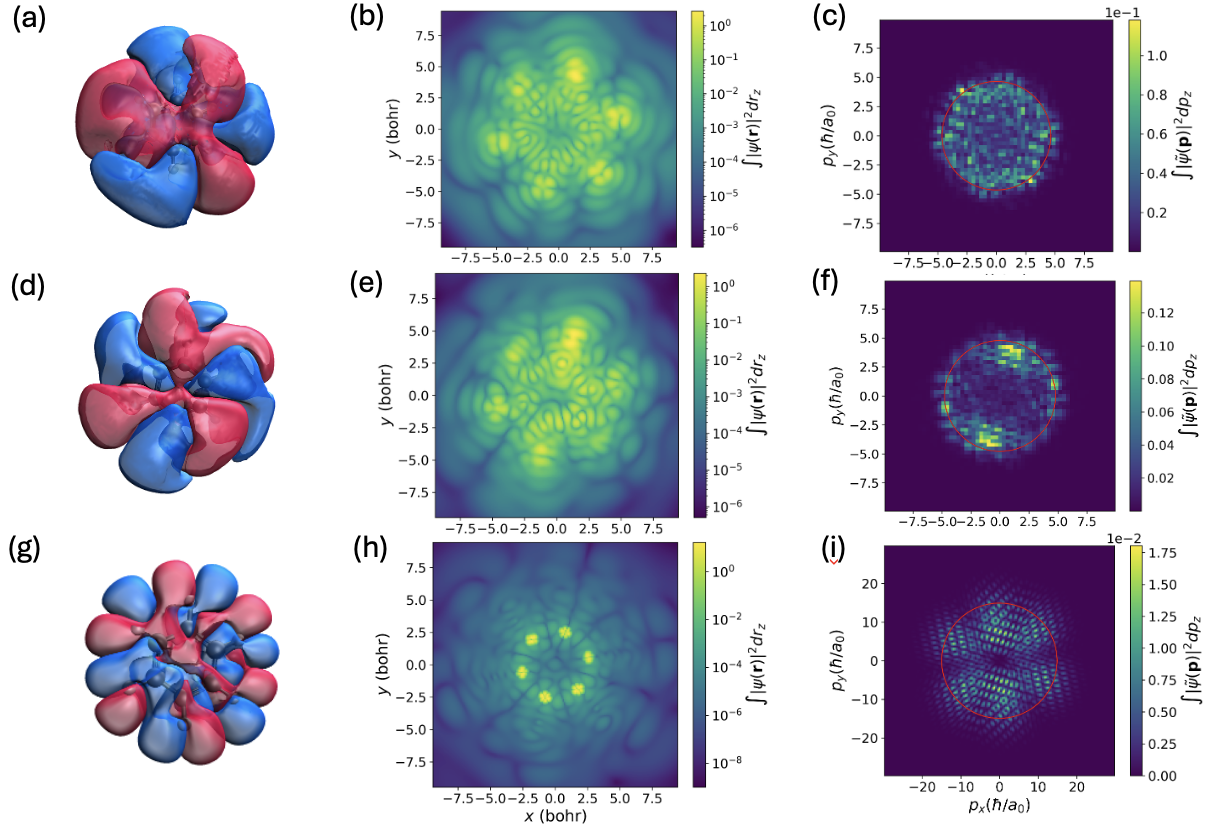}
    \caption{Three representative eigenfunctions of the perturbed benzene ($n=697$ - top panel, $n=704$ - middle panel, and $n=861$ - lower panel). The left panel displays the isosurface plots (isovalue = 0.01 a.u.) of molecular virtual orbital wavefunction, with red and blue colors representing opposite signs. The middle panel displays the position space $\int|\psi(\bm r)|^2 dr_z$ and the right panel is the momentum space projection onto the xy-plane $\int|\tilde{\psi}(\bm p)|^2dp_z$. The red circle plotted in the right panel represents equimomentum contour defined by the average kinetic energy of the orbital $p^2_x+p^2_y = 2m_e(\epsilon-\langle\hat{V}_{eff}\rangle)$.}
    \label{fig:wf}
\end{figure}

\subsection*{Random wave conjecture} We now examine the structure of single-particle chaotic wave-functions. In Ref.~\cite{BerryC}, Berry proposed an appealing heuristic microcanonical picture  of the Wigner function, $W({\bf r}, {\bf p}) = \int \psi^*({\bf r} + {\bm \rho}/2) \psi({\bm r} - {\bm \rho}/2) e^{i {\bf p}\cdot {\bm \rho}} d^3 \rho$, which was conjectured to be equidistributed on the classical manifold, $\epsilon=H({\bf r}, {\bf p})$. 

For chaotic billiards~\cite{Bunimovich,BunimovichGalitski}, this leads to a ring-shaped momentum space structure of the high-energy states. The molecular problem is formulated in non-compact space, includes both discrete and continuum states and hence is qualitatively different from billiards. To our knowledge there is no widely accepted theory of random eigenfunctions in this regime, which should be informed by the structure of classical electron trajectories in the self-consistent Hartree or Hartree-Fock potential. If the latter changes strongly around a classically chaotic orbit, we expect the Berry spherical momentum-shell to be smeared out. We found this to be the case for most excited states of the small molecules we studied. See, Fig.~\ref{fig:wf}, where  we observe regular structures, some of which may represent quantum scars~\cite{Scars}. Ring-shaped structures consistent with Berry's random wave conjecture can also be seen in the projected momentum-space wave-functions of select high-energy states. However these data should be interpreted with care, because the pseudocontinuum states may mimic random waves.

\begin{figure*}[t]
\centering
\includegraphics[width=\textwidth]
{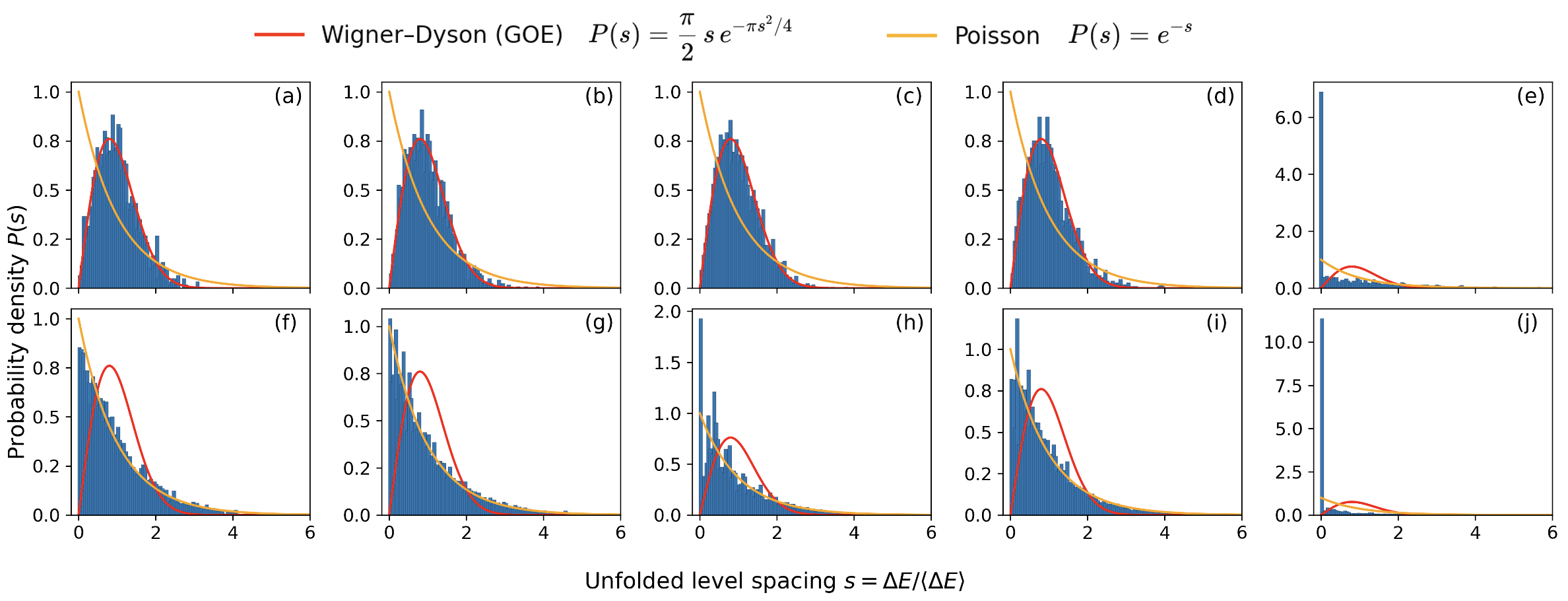}
    \caption{Nearest-neighbor level spacing statistics for (a)-(e) interacting and (f)-(j) non-interacting many particle excitations of (a)(f) methyloxirane, (b)(g)alanine,  (c)(h) 1-phenylethylamine, (d)(i) \(C_{1} \)  benzene , and (e)(j) \(D_{6h} \)   benzene.  The three chiral molecules and the perturbed benzene molecule have no spatial symmetries and the nearest-neighbor level spacings show GOE  distributions. The symmetric benzene molecule is known to have \(D_{6h} \) group symmetry and mixing excited states across symmetry groups disrupt the GOE distribution. }
    \label{fig:GOE}
\label{fig:wide}
\end{figure*}

\subsection*{Many-particle quantum chaos} In real molecules,  the energies of single-electron excitations are not the integrals of motion and only the many-body spectrum constrained by the total energy conservation is meaningful. To observe the emergence of many-body quantum chaos, we solve the electronic Hamiltonian in Eq.~\ref{eq:Hel} within the subspace of singly excited configurations built on the HF ground state, 
$
    \ket{\Psi^{a}_{i}} = \hat{a}^{\dagger}_a \hat{a}_{i}\ket{\Psi_0},
    $
 where $\hat{a}_{i}$ and $\hat{a}^{\dagger}_a $ denote the  annihilation and creation operators that act on the occupied orbital $i$ and the virtual orbital $a$. 
The excited wavefunction is then approximated as a linear combination of these singly excited configurations   
\begin{align}
    \ket{\Psi_J} & = \sum_{i,a} X^{J}_{ia} \ket{\Psi^{a}_{i}},
\end{align}
where the excitation amplitudes $X^{J}_{ia}$ are solved by diagonalizing the electronic Hamiltonian in Eq.~\ref{eq:Hel} in the single-excitation basis,
\begin{align}
    \sum_{j,b} H_{ia,jb}X^{J}_{jb} = E_{J}X^{J}_{ia}, \,\,\, 
    H_{ia,jb} = (\epsilon_a - \epsilon_i) \delta_{ij}\delta_{ab} + \langle ai||bj\rangle.\label{eq:CIS2}
\end{align}
The Coulomb and exchange two-electron integrals are 
\begin{align}
    \langle ai||bj\rangle &= \langle ai|bj\rangle - \langle ai|jb\rangle \\
    \langle pq|rs\rangle &= \int d\bm r_1 d\bm r_2 \frac{\psi^{*}_p(\bm r_1)\psi^{*}_q(\bm r_2)\psi_r(\bm r_1)\psi_s(\bm r_2)}{|\bm r_1 - \bm r_2|}.
\end{align}
Here indices $ij$ label occupied orbitals, $ab$ label virtual orbitals, and $pqrs$ denote general orbital indices. This corresponds to the standard configuration interaction singles (CIS) approximation in quantum chemistry~\cite{dreuw_single-reference_2005}.

If we ignore  interactions  for excited electron states by setting $\langle ai||bj\rangle=0$, there are no thermalization processes and each electron's energy is conserved. The model is chaotic in the single-particle sense, but integrable in the many-body sense~\cite{LiaoVikramGalitski}. We compute Slater determinants of the single-particle orbitals and observe  the Poisson statistics of the {\em many-body energies} as expected, see Fig.~\ref{fig:wide}~(f)--(j). 

Including interactions in earnest effectively makes the CIS Hamiltonian $ H_{ia,jb}$  a random matrix. Our numerical analysis of the {\em many-body spectrum} shows its excellent convergence to Wigner-Dyson GOE statistics, Figs.~\ref{fig:GOE}, for all molecules apart from \(D_{6h} \)  benzene, which shows an accumulation of low-energy states, see Fig.~\ref{fig:GOE}e. As before, this is due to high symmetry of benzene, whose  breaking restores Wigner-Dyson statistics - Fig.~\ref{fig:GOE}d. 

While, we have not examined the structure of chaotic many-electron eigenstates in detail, the eigenstate thermalization hypothesis (ETH)~\cite{ETH1,ETH2,MarcosTolya} predicts that they should  behave as pseudo-random vectors in the configuration basis. When expanded in a $D$-dimensional basis $\{\ket{\Phi_i}\}$, the coefficients $c_i^{J}$ of an eigenstate $\ket{\Psi_J} = \sum_{i=1}^{D} c_i^{J} \ket{\Phi_i}$ are expected to have zero mean and variance $\sim 1/D$. 

\subsection*{Spectral form-factor and quantum dynamics}
The spectral form-factor (SFF) is a more refined measure of level statistics than the level spacing histogram and is defined as a Fourier transform of the two-point level correlation function or equivalently~\cite{Mehta,Prange,Bound}
\begin{equation}
    \label{SFF}
    K(t) = \frac{1}{({\rm Tr}\, \hat{1})^2}  \left\langle | {\rm Tr}\, e^{-i\hat{H}t} |^2 \right\rangle
    \equiv \frac{1}{D^2} \sum\limits_{n,m=1}^D  \left\langle e^{-i (E_n - E_m) t} \right\rangle,
\end{equation}
where $D$ is the size of the relevant Hilbert space. In a finite-dimensional Hilbert space, the choice of a basis is irrelevant. Otherwise, including our case, it is dictated by physics. For example, if the molecule is kicked into an excited state spanned over a certain energy window, the latter determines the relevant subspace~\cite{Reimann}. 

 \begin{figure}[t]
\centering\includegraphics[width=1\linewidth]{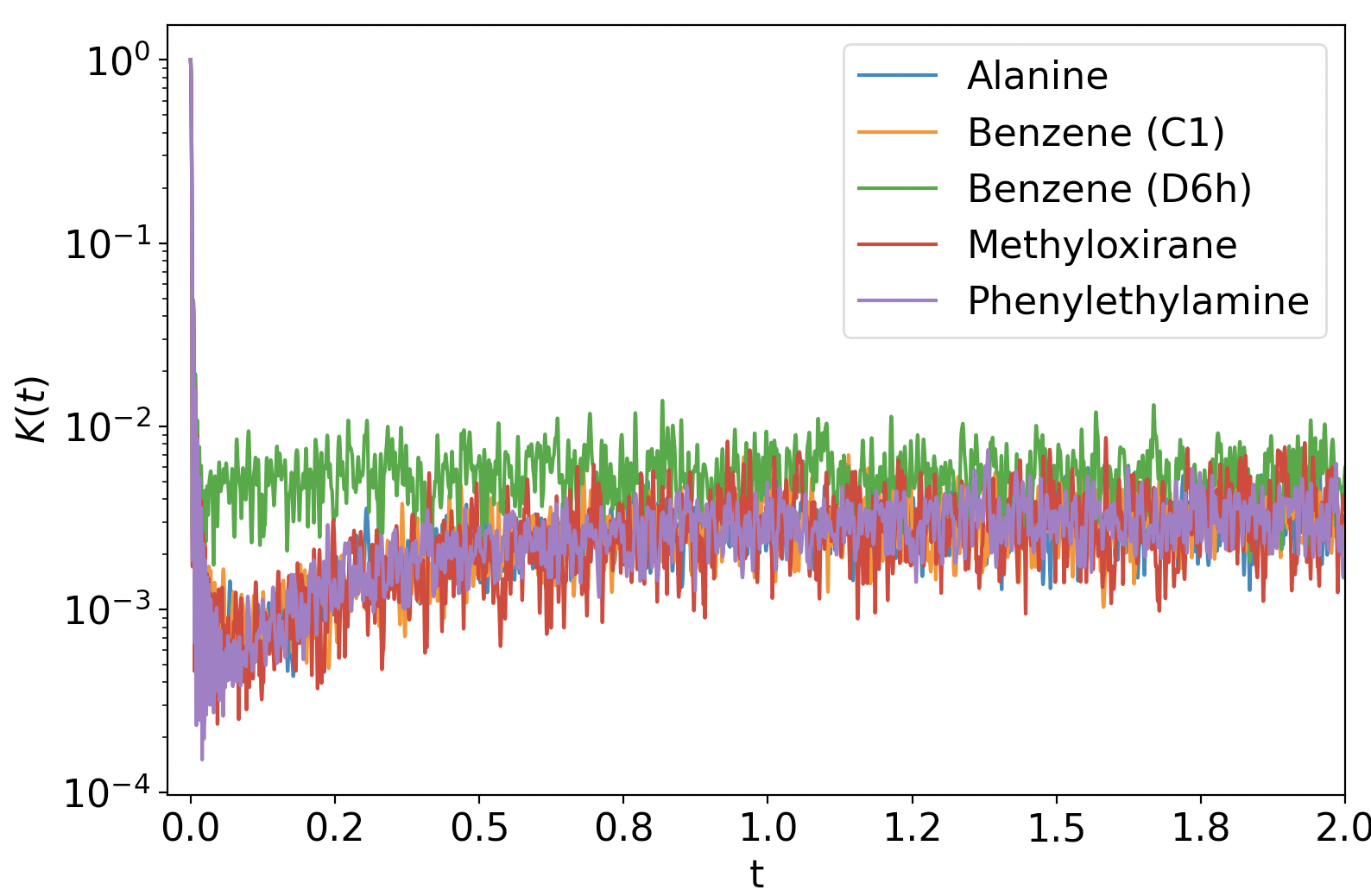}
    \caption{Spectral form factors calculated with unfolded energy levels for alanine (blue),  \(C_{1}\)  benzene (orange), \(D_{6h}\)  benzene (green), methyloxirane (red), 1-phenylethylamine (purple). The molecules that exhibit WD nearest-level statistics (all except  \(D_{6h}\)  benzene, shown in green) also display a ramp following the initial dip, consistent with a hallmark signature of quantum chaos.
    }
    \label{fig:sff}
\end{figure}

Fig.~\ref{fig:sff} displays the molecular SFFs, which collapse into the canonical ``quantum chaotic'' structure~\cite{BerryRigidity,Dephasing} for all geometries except the symmetric benzene: see, the dip, ramp, and plateau regions. At $t=0$, the SFF is trivially $K(0)=1$.  The early-time dynamics involves a sharp drop to $\sim 1/D$ - the dip - which occurs at the Thouless time. This perturbative regime is determined by the non-universal density of states (DoS), which is highly system specific and is {\em not} expected to follow the Wigner semicircle law~\cite{SpeedLimit}. 

Interestingly, the time it takes for the system to ``scramble'' (or ``forget'' the initial state, see Refs.~\cite{Reimann,Bound,SpeedLimit} for details) is approximately the Thouless time. While solving the time-dependent Schr{\"o}dinger equation is often a prohibitively difficult task, using the RMT approach circumvents the need to rederive statistical mechanics from scratch for each individual system, and ties quantum dynamics to relaively simple non-universal features of the bare SFF. 

The initial dip is followed by a ramp, which is the  hallmark feature of quantum chaos related to interference of unstable periodic orbits and time-translation symmetry~\cite{BerryRigidity,Gutzwiller,POrbits}. The ramp rises to a plateau value at the Heisenberg time and the SFF stays there. It corresponds to energy scales below the inverse level spacing where correlations are absent. We observe  the standard dip-ramp-plateau structure, which is accompanied by wild fluctuations representing finite-size effects. We emphasize that the ramp is a stronger evidence of quantum chaos than nearest-level repulsion, as it probes a more general pair correlation function. The latter appears in the distributions of unrelated structures from eigenenergies of billiards and disordered metals~\cite{AltlandKamenev} to non-trivial zeroes of the Riemann zeta function~\cite{Montgomery}.


\subsection*{Effect of the magnetic field} Next we consider level statistics in the presence of an external magnetic field by including it in Eq.~\ref{eq:Hel} via minimal substitution
$ \hat{\bm p }_i \rightarrow \hat{\bm p }_i+e\bm A( \hat{\bm r}_i)$, where $\bm A( \hat{\bm r}_i) = \frac{1}{2}\bm B\times(\hat{\bm r}_i - \bm G)$ is the vector potential and $\hat{\bm r}_i$ is the position operator of the $i$th-electron. In quantum chemistry, it is customary to include the ``gauge origin,'' $\bm G$, even though the additional term represents a pure gauge. Obviously, any gauge transform including a shift of ${\bf G}$ must have no effect on observables.  In a finite computational basis of Gaussian atomic orbitals however, ensuring gauge invariance is non-trivial as discussed in Refs.~\cite{london_theorie_1937,helgaker_electronic_1991}.  We ensured gauge invariance, including  $\frac{\partial E}{\partial \bm G}=  0$, by employing the standard procedure in quantum chemistry based on the London orbitals~\cite{london_theorie_1937,helgaker_electronic_1991}. This entails attaching an atomic-centered $U(1)$ gauge phase-factor to each atomic  basis function; i.e., ${\phi}_{\rm London}(\bm r) \equiv e^{-\frac{i}{2\hbar}\bm B\times(\bm R_M - \bm G)\cdot \bm r}\phi(\bm r-\bm R_M)$, where $M$ labels the nuclear center of the basis function.

\begin{figure}[t]
\centering\includegraphics[width=1\linewidth]{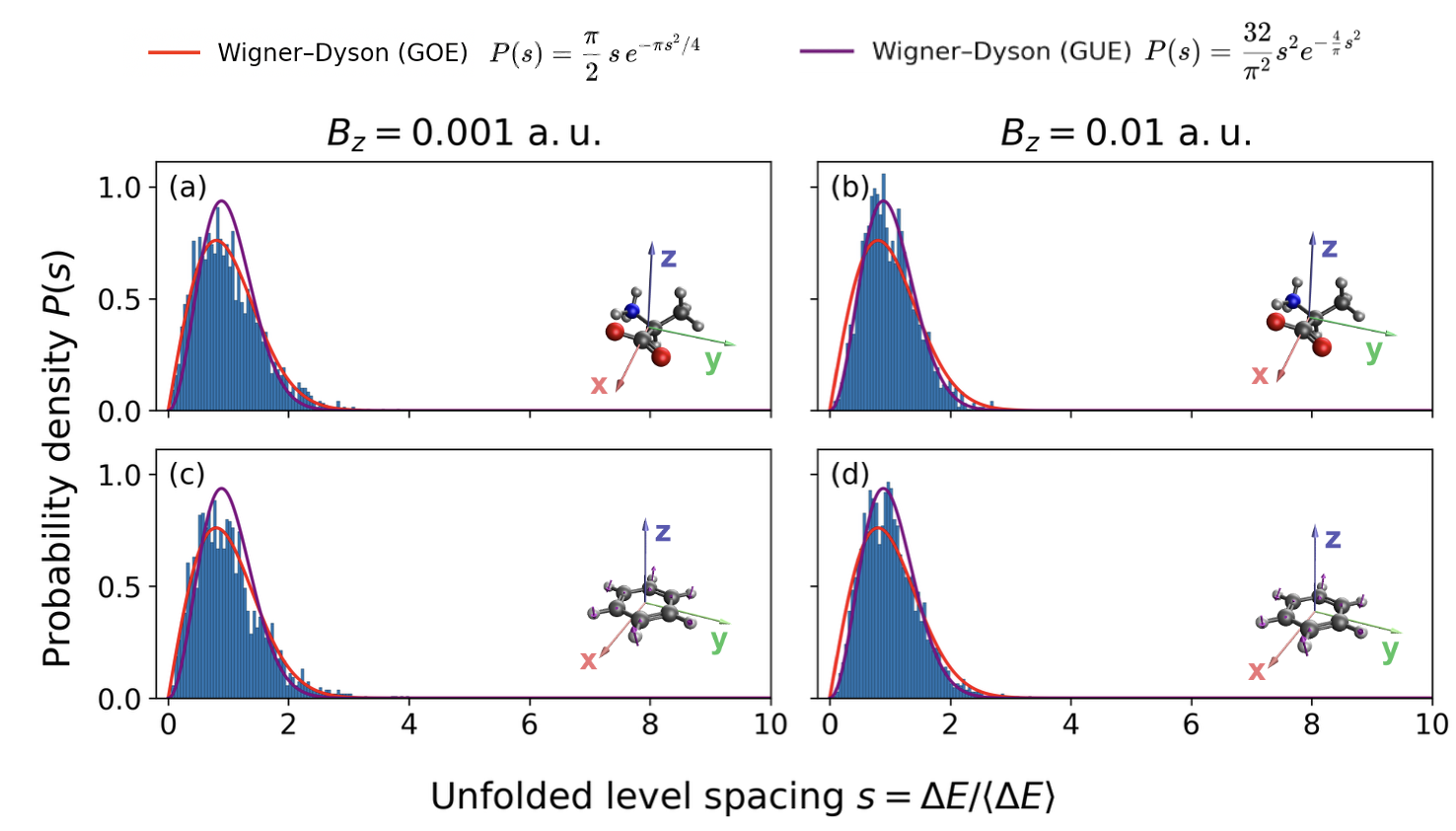}
    \caption{Nearest-neighbor level spacing statistics for interacting CIS  excitations of (a)-(b) alanine, (c)-(d) perturbed benzene in an external magnetic field along the z direction with $B_z = 0.001$ a.u. in (a)(c) and $B_z = 0.01$ a.u. in (b)(d). The x, y, and z axes are shown in the molecular frames in orange, green, and blue. The transition from GOE to GUE becomes more apparent with $B_{z}=10^{-2}$ a.u..}
    \label{fig:GUE} 
\end{figure}

The magnetic  field  breaks time-reversal symmetry and is expected to change the WD statistics from the Gaussian orthogonal to Gaussian unitary ensemble (GUE) in the absence of other antiunitary symmetries (as present for magnetic hydrogen~\cite{MHydrogen} and we also expect this symmetry asymptotically restored for highly excited molecular Rydbergs). However, capturing the GUE is non-trivial for small molecules because the magnetic energy scale is often smaller than the relevant level spacing. Fig. \ref{fig:GUE}  plots the nearest-neighbor many-body level spacing statistics for alanine and  \(C_{1}\)  benzene molecules where the statistics remains GOE for physically meaningful out-of-plane field.    To enforce the transition to GUE in simulations, we applied an  extremely large magnetic field, $B_{z}>10^{-2}$a.u. This scale can be qualitatively understood by comparing the flux through an electronic orbital $\Phi \sim \pi B_z \langle \hat{ r}^2\rangle$ to the flux quantum.  

This however does not mean that GUE is out of reach for realistic electronic spectra. Spectra of larger molecular structures (e.g., multiple-ring  helicene) are expected to turn GUE for smaller orbital fields. 
Furthermore,  application of a circularly polarized light may mimic the effect of an orbital magnetic field~\cite{Gedik,FTI}, albeit in non-equilibrium with ${\bf A}(t) = F_0/\Omega \left(\pm \cos{\Omega t},\sin{\Omega t}, 0\right)$, where $F_0$ is the amplitude of the periodic electric field, ${\bf F}(t) = - \partial_t {\bf A}$ and $\Omega=2\pi/T$ is its frequency. The solution to the time-dependent Schr{\"o}dinger equation for the time-evolution operator, $i \partial_t \hat{U}(t) = \hat{H}(t) \hat{U}(t)$ satisfies the Floquet theorem~\cite{Floquet}, $\hat{U}(t) = \hat{P}(t) e^{-i\hat{\cal F} t}$, where $ \hat{P}(t)=  \hat{P}(t+T)$ is a periodic unitary which resets every period, $ \hat{P}(nT) = \hat{1}$ and  $\hat{\cal F}$ is the {\em time-independent} Floquet Hamiltonian, which determines effective electron properties~\cite{SubotnikFloquet,Gedik,FTI}. Since the spectrum of $\hat{\cal F}$ is defined on a circle modulo $\Omega$,  we expect the Floquet unitary operator to crossover from the circular orthogonal ensemble (COE) to the circular unitary ensemble  (CUE) as the drive strength increases.

\begin{figure}[t]
\centering\includegraphics[width=1\linewidth]{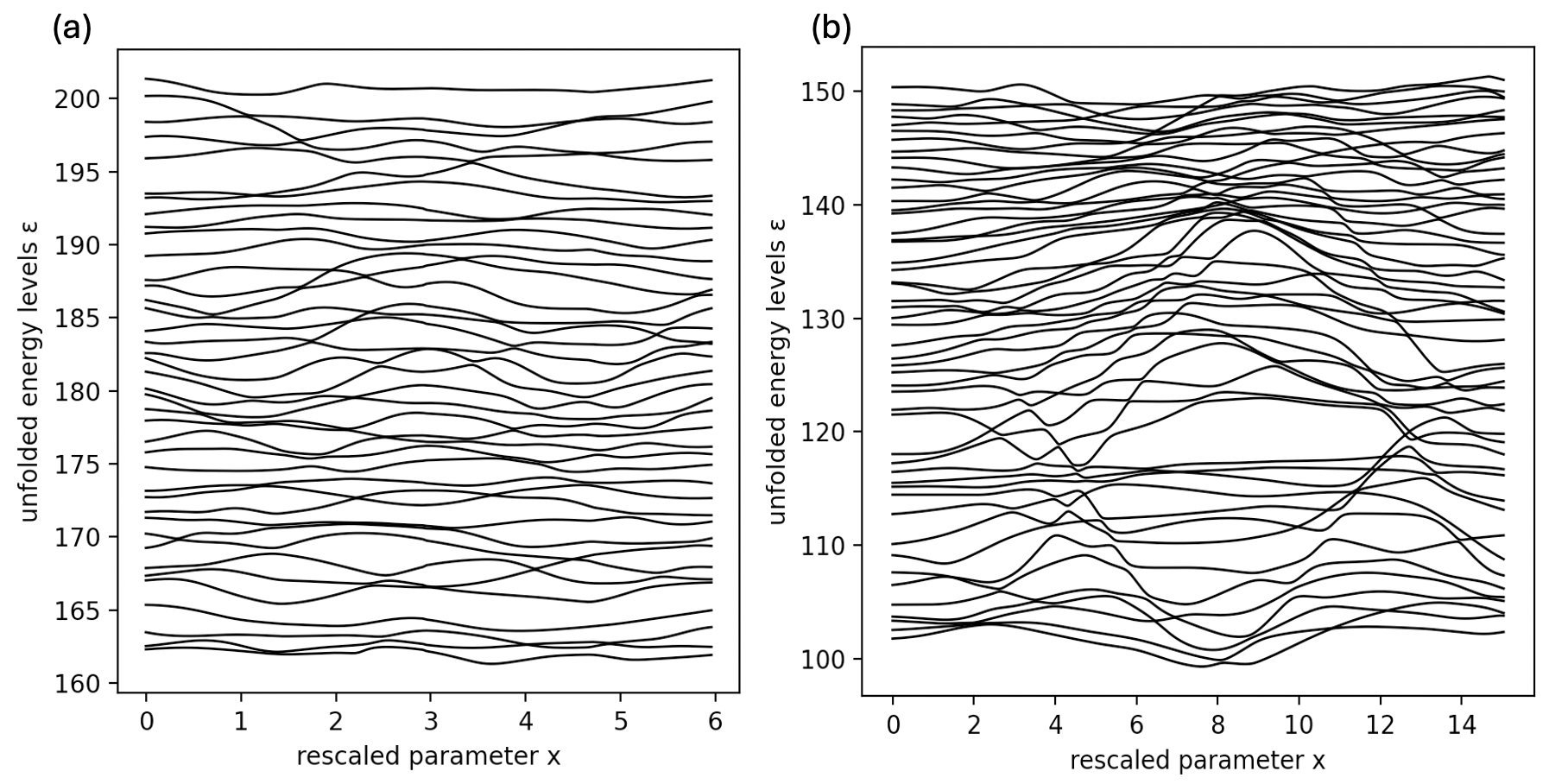}
    \caption{Unfolded energies $\epsilon_I = \frac{E_I}{\Delta}$ with a mean level spacing $\Delta$ as a function of the unfolded parameter (a) $x = \sqrt{C_0} (F- F_0)$, where $C_0 = \langle\Big(\frac{\partial \epsilon}{\partial F}\Big)^2\rangle$ is the averaged level velocity variance and  $F$ is the external electric field. (b) $x = \sqrt{C_0}t$, where $C_0 = \langle\Big(\frac{\partial \epsilon}{\partial t}\Big)^2\rangle$ is the averaged level velocity variance. The energy levels are varied as a function of time, $t$, due to atomic motion ${\bm R}_a(t)$.
    }
    \label{fig:noodle}
\end{figure} 

\subsection*{Level velocities and curvatures} Changing the Hamiltonian in Eq.~\ref{eq:Hel} by either displacing the atoms due to  atomic motion, ${\bf R}(t)$,~\footnote{We assume a separation of electronic and nuclear time scales, so that vibrational motion is much slower
than electronic dynamics. The electronic spectrum is treated as an
instantaneous function of the nuclear coordinates.
Non-equilibrium effects and the electromagnetic
fields induced by the resulting time-dependent currents (which could 
break time-reversal symmetry) are neglected.} or the application of the electric field  changes the realization of the electronic spectrum -- energy levels move.  Specifically, for the electric field, ${\bf F}$, the Hamiltonian reads. 
\begin{align}
    \hat{H}_{\rm F}(\bm r, \bm R,\bm F) =  \hat{H}_{\rm el}(\bm r, \bm R) -  \hat{\bm \mu}_{e} \cdot \bm F.
    \label{F}
\end{align}
 Using perturbation theory,  its eigenstates $\{\psi\}$   can be written in terms of the eigenstates $\{\phi\}$  of the unperturbed Hamiltonian, $\ket{\psi_I(\bm F)}
    = \ket{\phi_I}
    + \sum_{J \neq I}
      \frac{\bra{\phi_J}\,\hat{\bm\mu}_e \cdot \bm F\,\ket{\phi_I}}
           {E_J - E_I}\,\ket{\phi_J}
    + \mathcal{O}(|\bm F|^2)$. Hence, the level velocity  for the state $I$ with respect to the field (corresponding to the electric dipole matrix element)  is
$
    \frac{\partial E_I} {\partial \bm F} =  - \bra{\psi_I}   \bm   \hat{\bm \mu}_{e}  \ket{\psi_I} 
$
by the Hellmann-Feynman theorem. The level curvature (the diagonal components of the electric polarizability tensor) is
\begin{align}
    \frac{\partial^2 E_I} {\partial \bm F^2} =
    & = 2\sum_{J\ne I } \frac{|\bra{\psi_J}\hat{\bm \mu}_{e}\ket{\psi_I}|^2}{E_J-E_I} \label{eq:curvature}
\end{align}

Fig.~\ref{fig:noodle}a shows the flow of random matrix spectra as a function of the electric field. Fig.~\ref{fig:noodle}b  is the same but under nuclear motion 
 parameterized by time, $t$, where the nuclei trace out a realistic trajectory, ${\bf R}(t)$ calculated using molecular dynamics simulations.  In both cases, we observe a highly ``chaotic'' structure, which illustrates the sensitivity of electron spectra and suggests that Dyson Brownian motion~\cite{DysonBM} due to environmental noise may render individual highly excited electronic levels ill-defined in practice for complex molecules and materials.
 
\begin{figure}[t]
\centering\includegraphics[width=1\linewidth]{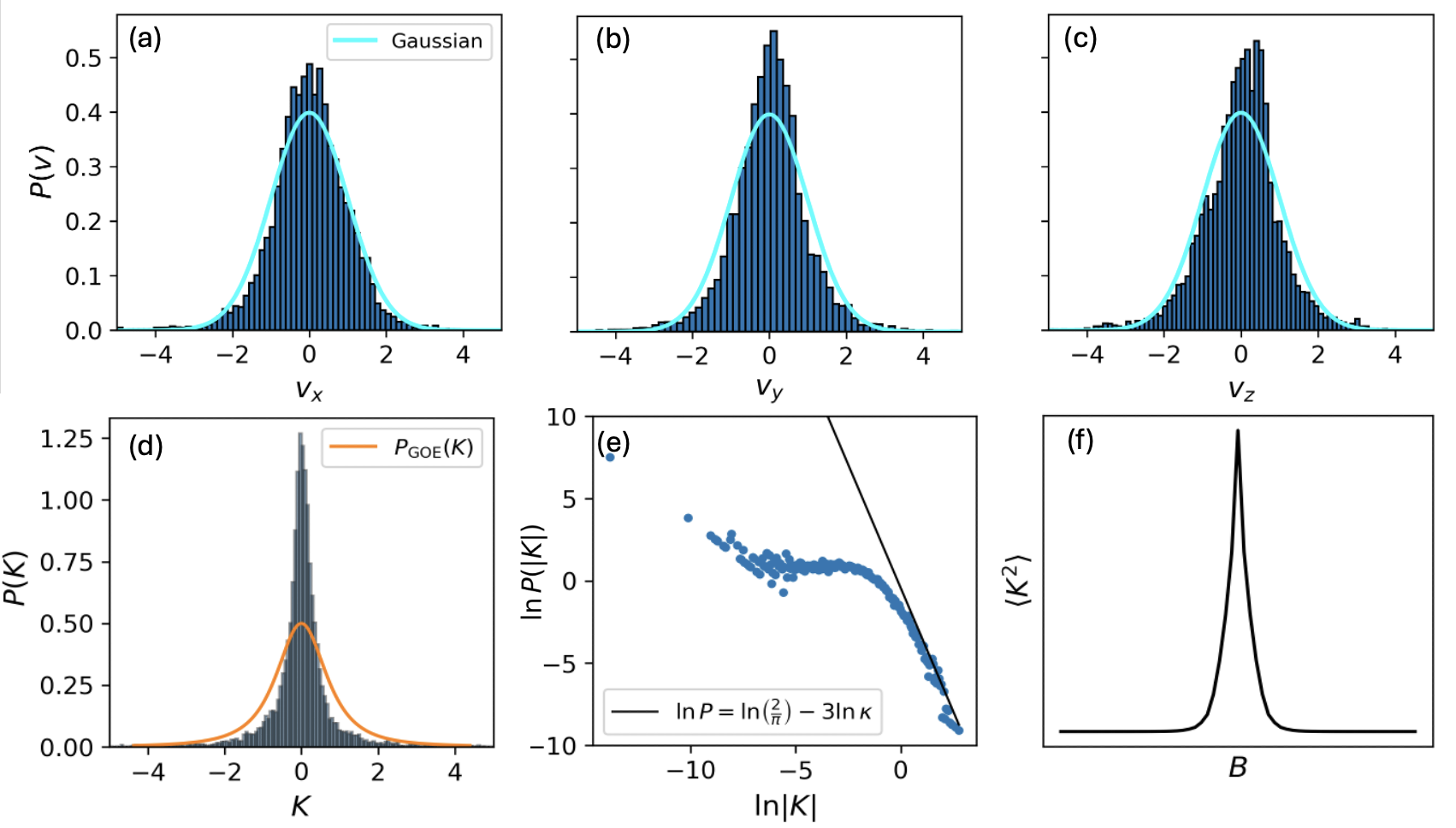}
    \caption{Level velocity and curvature statistics of the \(C_{1} \)    benzene in a varying external electric field $\bm F$. Histograms of unfolded level velocities $v_{\sigma}=\frac{\partial E}{\partial F_{\sigma}} /\sqrt{\langle(\frac{\partial E}{\partial F_{\sigma}})^2\rangle}$ along $\sigma = x, y, z$ axis are shown in (a)-(c), respectively, which approximates the transition electric dipole moment within the CIS framework. The  histogram and tail statistics for the unfolded isotropic curvature $\kappa = \frac{1}{3}\sum_{\sigma}\frac{\partial^2 E}{\partial F_{\sigma}^2} /\langle(\frac{1}{3}\sum_{\sigma}\frac{\partial E}{\partial F_{\sigma}})^2\rangle$ is shown in (d) and (e), which approximates the isotropic polarizability within the CIS framework. The variance of the curvature diverges at $\bm B \rightarrow 0 $ is shown schematically in  (f). }
    \label{fig:curve}
\end{figure}

 However, this level dynamics has a hidden statistical structure~\cite{SimonsAltshuler,ZD1,ZD2,vonOppen}. For the canonical Gaussian random matrix ensemble, the distribution of level velocities is Gaussian. However, this Gaussianity is non-universal being sensitive to the scar structure of more realistic physical systems. As a representative example, we show the velocity statistics of the electronic spectra of perturbed benzene in Fig.~\ref{fig:curve}(a-c), where small velocity structure does differ from a Gaussian. Within the GOE RMT, the full distribution of level curvatures for unfolded eigenvalues was conjectured by
Zakrzewski and Delande~\cite{ZD1} to take the  form
\begin{equation}
    P_{\mathrm{GOE}}(K)
    = \frac{1}{2\gamma_0}\,
      \frac{1}{\bigl[1+(K/\gamma_0)^2\bigr]^{3/2}},
    \label{eq:ZD_PofK}
\end{equation}
where $\gamma_0$ is a curvature scale fixed by the mean level spacing and the variance of the level velocities. The low-$K$ regime corresponds to an adiabatic level dynamics, while the universal high-$K$ limit~\cite{vonOppen,FyodorovSommers} is related to the statistics of avoided level-crossings. They are shadows of the nearby conical intersections~\cite{Con1,Con2} (or equivalently ``diabolical points''~\cite{Diablo}) accessible through two-parameter flows or possibly higher-dimensional degeneracy structures in the larger parameter space~\cite{KitaevTI}. However, the exact degeneracies have zero probability to occur within a one-parameter flow without special symmetries or fine-tuning. This is the case in our simulations, Fig.~\ref{fig:curve}, where we recover the universal  $P(K)\propto |K|^{-3}$ tail.

Because of the slow $1/|K|^3$ decay, the second moment of $P_{\mathrm{GOE}}(K)$ diverges logarithmically.  The infrared regularization is dictated by physics of the problem. For example a weak field, ${\bf B} = B {\bf e}_z$ introduces off-diagonal terms in the  minimal $2\times2$ matrix model of the avoided crossing and cuts off the variance of the curvature:
\begin{equation}
    \bigl\langle K^2 \bigr\rangle \sim
    \gamma_0^2 \ln\!\left(\frac{B_*}{|B|}\right),
    \qquad B\to 0.
    \label{eq:K2_B}
\end{equation}
where $B_*$ is a constant. \eqref{eq:K2_B} shows that the ensemble-averaged curvature in the orthogonal regime grows logarithmically as $B\to 0$. This is reminiscent of the weak localization correction to conductivity, which scales the same way as Eq.~\ref{eq:K2_B} in two dimensional systems, and is ubiquitously observed in experiment~\cite{AA}.

\subsection*{Character of the electronic RMT states}  
\begin{figure}[t]
\centering\includegraphics[width=1\linewidth]{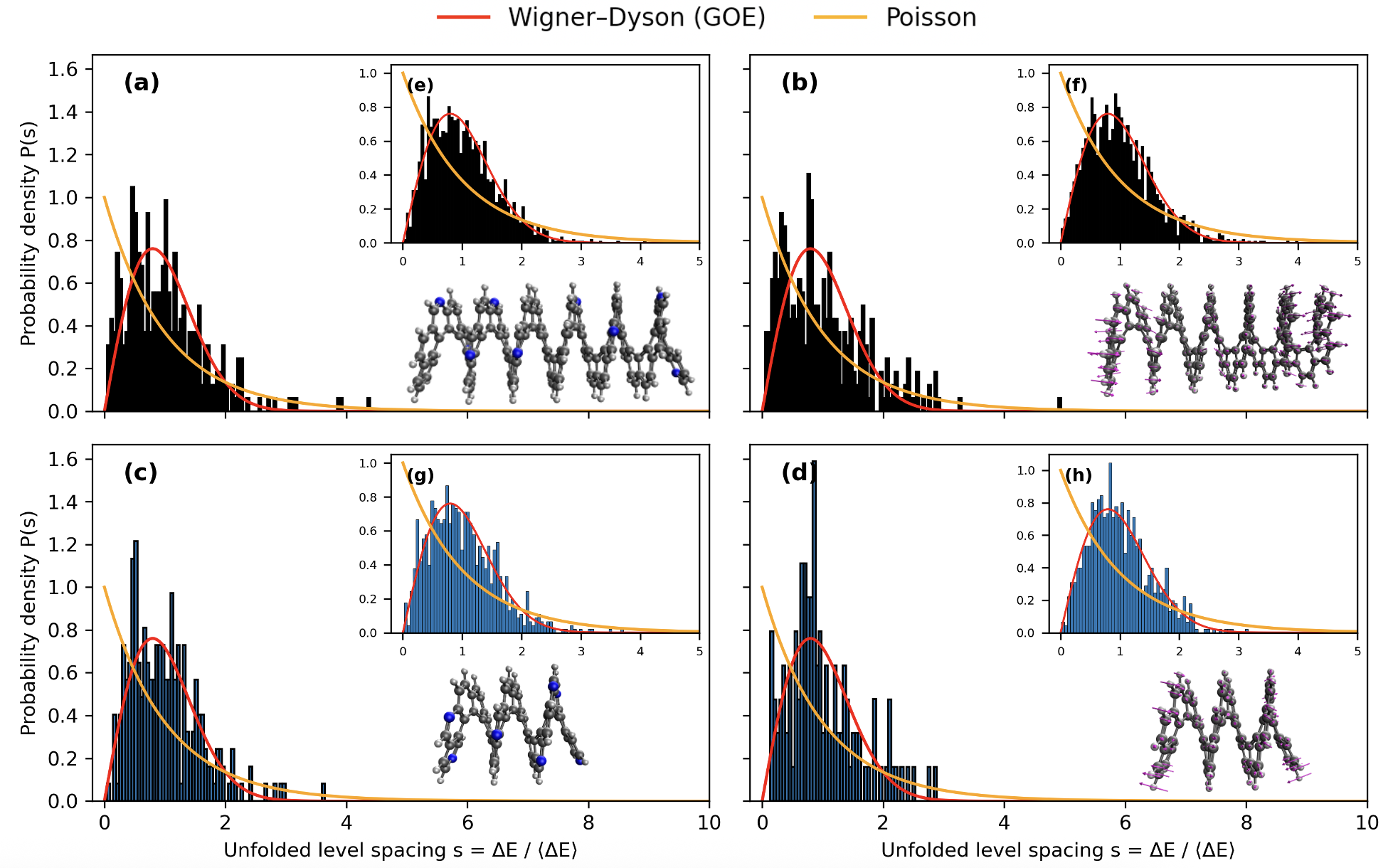}
    \caption{Nearest-neighbor unfolded level spacing statistics for (a)-(b) bound states and (e)-(f) full spectra of non-interacting orbital energies (black) of [36]-helicene and (c)-(d) bound states and (g)-(h) both bound and unbound states of interacting  excited states (blue) of [18]-helicene. The $C_2$ symmetry of these helicene chains is broken either through substitution of nitrogen atoms as shown in (a)(c)(e)(g) or displacement of nuclear Cartesian coordinates along the 6th-vibrational mode as shown in (b)(f)(d)(h). The histograms constructed from both bound and unbound states exhibit WD GOE statistics. Similar characteristics is observed when only the bound states are considered, with the degree of agreement depending on the extent to which the $C_2$ symmetry is broken. }
    \label{fig:bound}
\end{figure}

In the single-particle description, the electronic states can be classified in increasing energy as core orbitals, valence orbitals, valence virtual orbitals, Rydberg orbitals, and continuum states. The continuum states have no energy gaps and hence have no simple notion of level statistics in terms of discrete level spacings. Yet, the finite Gaussian atomic orbital basis effectively introduces pseudocontinuum states as discretized artifacts in the simulations.  The Rydberg orbitals are hydrogen-like and there are an infinite number of them (they should still yield WD statistics in the presence of strong interactions due to multi-electron excitations, but observing it in finite-basis simulations is challenging). To connect imperfect simulations to the spectra of real molecules, it is therefore important to remove the pseudocontinuum and Rydberg states and isolate the valence orbitals which tend to couple strongly. To collect meaningful statistics of these discrete states, we study long-chain helicene molecules with broken $C_2$ symmetry through atomic substitution and atomic displacement along the $C_2$-symmetry-breaking vibrational mode. In Fig. \ref{fig:bound}, we used density functional theory (DFT) method to plot the neighboring orbital energy differences for the bound states, excluding the localized core orbitals. Here, we still see clear tendency towards the WD statistics which is a more quantitatively reliable result.
 
 We also perform a study of the many-body level statistics of the valence states (by explicitly excluding the  Rydberg orbitals, core excitations, resonances and continuum states for the long helicene molecules). Similarly, we observe many-body quantum chaos generated by these bound states. Fig.~\ref{fig:bound} plots the neighboring level statistics for the excited states below the first ionization threshold calculated using linear-response time-dependent DFT within the Tamm-Dancoff Approximation (LR-TDDFT/TDA). ~\cite{hirata_time-dependent_1999} We performed all the calculations in a developmental branch of Q-Chem software~\cite{Epifanovsky2021}. 


\section*{Conclusions}
We have shown that ab initio electronic structure methods effectively generate random matrices. Although we restrict our analysis to the singlet single excitations, the lower-energy bound eigenstates of these matrices can provide accurate predictions of the molecular energetics.  In finite basis sets, high-lying states are not necessarily individually accurate, but their statistical properties are expected to be robust, universal, and experimentally testable. This conceivably may  explain why increasing the basis size and summing over highly excited electron levels often improves convergence of quantum chemistry computations and their correspondence to experimental data~\cite{dalessandro_locality_2019,otero_new_2019}. In both cases, there may be two key contributions at play:  low-energy non-chaotic levels (captured accurately by ab initio numerics) and high-energy RMT levels. To verify these conjectures, we propose measuring universal statistics of level crossings through electric and magnetic polarizabilities~\cite{hohm_experimental_2013}. Experimentally probing the Wigner-Dyson statistics of the molecular electronic structure directly should also be within reach. Even though separating electronic excitations from vibronic states may be difficult in practice, the combined spectrum may still exhibit RMT-like statistics in regimes of strong mixing. Finally, beyond the bulk Wigner-Dyson statistics studied here, RMT also predicts universal behavior at the spectral edge, where the fluctuations of the  smallest eigenvalue follow the Tracy-Widom distribution~\cite{TW1,TW2}. 

This work raises follow-up questions about the signatures of quantum chaos arising from different physical interactions in chemical systems. On the one hand, stronger electronic correlation can be further investigated using the state-of-the-art electronic structure methods that account for higher-order excitations and multireference character~\cite{gonzalez_progress_2012,lischka_multireference_2018}. On the other hand, vibronic mixing becomes increasingly important for highly excited electronic states in regimes of large density of states, which can give rise to chaotic features beyond the Born-Oppenheimer picture~\cite{Abedi2012,agostini_chemistry_2022,bian_phase-space_2026}. Finally, although spin-orbit coupling is generally weak in organic molecules and we neglected it here, stronger spin-orbit-coupled molecules are expected to exhibit features of the third canonical RMT ensemble - Gaussian symplectic ensemble (GSE) - with Kramers-degenerate levels and a distinct pattern of level repulsion in molecular spectra.

\acknowledgements
This work was supported by startup funds from the University of Rhode Island (Z.T.), the U.S. Department of Energy, Office of Science Basic Energy Sciences under Award No. DE-SC0001911 and the Julian Schwinger Foundation (V.G.). We acknowledge the use of the Unity Cluster for computational resources. Z.T. would like to thank Tanner Culpitt for helpful discussions and assistance with the magnetic field calculations.

\bibliography{Ref}

\begin{thebibliography}{78}
\providecommand{\natexlab}[1]{#1}
\providecommand{\url}[1]{\texttt{#1}}
\expandafter\ifx\csname urlstyle\endcsname\relax
  \providecommand{\doi}[1]{doi: #1}\else
  \providecommand{\doi}{doi: \begingroup \urlstyle{rm}\Url}\fi

\bibitem[Gutzwiller(1990)]{Gutzwiller}
M.~C. Gutzwiller.
\newblock \emph{Chaos in Classical and Quantum Mechanics}.
\newblock Springer, Berlin, 1990.

\bibitem[Poincar{\'e}(1892-1899)]{Poincare}
Henri Poincar{\'e}.
\newblock \emph{Les m{\'e}thodes nouvelles de la m{\'e}canique c{\'e}leste}.
\newblock Gauthier-Villars, Paris, 1892-1899.
\newblock 3 vols.: Tome I (1892), Tome II (1893), Tome III (1899).

\bibitem[Bohigas et~al.(1984)Bohigas, Giannoni, and Schmit]{BGS_1984}
O.~Bohigas, M.~J. Giannoni, and C.~Schmit.
\newblock Characterization of chaotic quantum spectra and universality of level
  fluctuation laws.
\newblock \emph{Phys. Rev. Lett.}, 52\penalty0 (1):\penalty0 1--4, 1984.
\newblock \doi{10.1103/PhysRevLett.52.1}.
\newblock URL \url{https://link.aps.org/doi/10.1103/PhysRevLett.52.1}.
\newblock Publisher: American Physical Society.

\bibitem[Mehta(2004)]{Mehta}
Madan~Lal Mehta.
\newblock \emph{Random matrices}.
\newblock Academic Press, Inc., Boston, MA, third edition, 2004.

\bibitem[Haake(2010)]{Haake}
Fritz Haake.
\newblock \emph{Quantum Signatures of Chaos}.
\newblock Springer-Verlag Berlin Heidelberg, 2010.

\bibitem[Dyson(1962)]{DysonBM}
Freeman~J. Dyson.
\newblock A {B}rownian-motion model for the eigenvalues of a random matrix.
\newblock \emph{Journal of Mathematical Physics}, 3\penalty0 (6):\penalty0
  1191--1198, 1962.
\newblock ISSN 0022-2488.
\newblock \doi{10.1063/1.1703862}.

\bibitem[Wigner(1951)]{Wigner}
Eugene~P. Wigner.
\newblock On the statistical distribution of the widths and spacings of nuclear
  resonance levels.
\newblock \emph{Mathematical Proceedings of the Cambridge Philosophical
  Society}, 47:\penalty0 790--798, 1951.

\bibitem[Beenakker(1997)]{Beenakker}
C.~W.~J. Beenakker.
\newblock Random-matrix theory of quantum transport.
\newblock \emph{Reviews of Modern Physics}, 69:\penalty0 731--808, 1997.
\newblock \doi{10.1103/RevModPhys.69.731}.

\bibitem[Altland and Zirnbauer(1996)]{AZ}
Alexander Altland and Martin~R. Zirnbauer.
\newblock Random matrix theory of a chaotic andreev quantum dot.
\newblock \emph{Physical Review Letters}, 76:\penalty0 3420--3423, 1996.
\newblock \doi{10.1103/PhysRevLett.76.3420}.

\bibitem[Wasserman et~al.(2008)Wasserman, Maitra, and
  Heller]{wasserman_investigating_2008}
Adam Wasserman, Neepa~T. Maitra, and Eric~J. Heller.
\newblock Investigating interaction-induced chaos using time-dependent
  density-functional theory.
\newblock \emph{Phys. Rev. A}, 77\penalty0 (4):\penalty0 042503, 2008.
\newblock \doi{10.1103/PhysRevA.77.042503}.
\newblock URL \url{https://link.aps.org/doi/10.1103/PhysRevA.77.042503}.
\newblock Publisher: American Physical Society.

\bibitem[Swingle et~al.(2016)Swingle, Bentsen, Schleier-Smith, and
  Hayden]{Swingle}
Brian Swingle, Gregory Bentsen, Monika Schleier-Smith, and Patrick Hayden.
\newblock Measuring the scrambling of quantum information.
\newblock \emph{Physical Review A}, 94:\penalty0 040302, 2016.
\newblock \doi{10.1103/PhysRevA.94.040302}.

\bibitem[{Google~Quantum~AI~and~Collaborators}(2021)]{Google}
{Google~Quantum~AI~and~Collaborators}.
\newblock Information scrambling in computationally complex quantum circuits.
\newblock \emph{Science}, 374:\penalty0 1479--1483, 2021.
\newblock \doi{10.1126/science.abg5029}.

\bibitem[Joshi et~al.(2022)Joshi, Elben, Vikram, Vermersch, Galitski, and
  Zoller]{GalitskiZoller}
Lata~Kh. Joshi, Andreas Elben, Amit Vikram, Beno{\^\i}t Vermersch, Victor
  Galitski, and Peter Zoller.
\newblock Probing many-body quantum chaos with quantum simulators.
\newblock \emph{Physical Review X}, 12:\penalty0 011018, 2022.
\newblock \doi{10.1103/PhysRevX.12.011018}.

\bibitem[Dong et~al.(2025)Dong, Zhang, Dag, Gao, Wang, Deng, Zhang, Chen, Xu,
  Wang, Wu, Zhang, Jin, Zhu, Zhang, Zou, Tan, Cui, Zhu, Shen, Li, Zhong, Bao,
  Li, Wang, Guo, Song, Liu, Chan, Ying, and Wang]{SFFexp}
Hang Dong, Pengfei Zhang, Ceren~B. Dag, Yu~Gao, Ning Wang, Jinfeng Deng,
  Xu~Zhang, Jiachen Chen, Shibo Xu, Ke~Wang, Yaozu Wu, Chuanyu Zhang, Feitong
  Jin, Xuhao Zhu, Aosai Zhang, Yiren Zou, Ziqi Tan, Zhengyi Cui, Zitian Zhu,
  Fanhao Shen, Tingting Li, Jiarun Zhong, Zehang Bao, Hekang Li, Zhen Wang,
  Qiujiang Guo, Chao Song, Fangli Liu, Amos Chan, Lei Ying, and H.~Wang.
\newblock Measuring spectral form factor in many-body chaotic and localized
  phases of quantum processors.
\newblock \emph{Physical Review Letters}, 134:\penalty0 010402, 2025.
\newblock \doi{10.1103/PhysRevLett.134.010402}.

\bibitem[Aaronson and Arkhipov(2013)]{AA}
Scott Aaronson and Alex Arkhipov.
\newblock The computational complexity of linear optics.
\newblock \emph{Theory of Computing}, 9:\penalty0 143--252, 2013.
\newblock \doi{10.4086/toc.2013.v009a004}.

\bibitem[Madsen et~al.(2022)Madsen, Laudenbach, Askarani, Rortais, Vincent,
  Bulmer, Miatto, Neuhaus, Helt, Collins, and et~al.]{Xanadu}
L.~S. Madsen, F.~Laudenbach, M.~F. Askarani, F.~Rortais, T.~Vincent, J.~F.
  Bulmer, F.~M. Miatto, L.~Neuhaus, L.~G. Helt, M.~J. Collins, and et~al.
\newblock Quantum computational advantage with a programmable photonic
  processor.
\newblock \emph{Nature}, 606:\penalty0 75--81, 2022.
\newblock \doi{10.1038/s41586-022-04725-x}.

\bibitem[Shou et~al.(2025)Shou, Miller, and Galitski]{Hiding}
Laura Shou, Sarah~H. Miller, and Victor Galitski.
\newblock Proof of hiding conjecture in gaussian boson sampling.
\newblock \emph{arXiv}, 2025.

\bibitem[Maldacena et~al.(2016)Maldacena, Shenker, and Stanford]{MSS}
Juan Maldacena, Stephen~H. Shenker, and Douglas Stanford.
\newblock A bound on chaos.
\newblock \emph{Journal of High Energy Physics}, 2016\penalty0 (8):\penalty0
  106, 2016.
\newblock \doi{10.1007/JHEP08(2016)106}.

\bibitem[Sachdev and Ye(1993)]{SY}
Subir Sachdev and Jinwu Ye.
\newblock Gapless spin-fluid ground state in a random quantum {Heisenberg}
  magnet.
\newblock \emph{Physical Review Letters}, 70:\penalty0 3339--3342, 1993.
\newblock \doi{10.1103/PhysRevLett.70.3339}.

\bibitem[Abramson et~al.(1984)Abramson, Field, Imre, Innes, and
  Kinsey]{abramson_stimulated_1984}
Evan Abramson, Robert~W. Field, Dan Imre, K.~K. Innes, and James~L. Kinsey.
\newblock Stimulated emission pumping of acetylene.
\newblock \emph{The Journal of Chemical Physics}, 80\penalty0 (6):\penalty0
  2298--2300, 1984.
\newblock ISSN 0021-9606.
\newblock \doi{10.1063/1.447006}.
\newblock URL \url{https://doi.org/10.1063/1.447006}.

\bibitem[Persch et~al.(1988)Persch, Mehdizadeh, Demtröder, Zimmermann,
  KÖppel, and Cederbaum]{persch_vibronic_1988}
G.~Persch, E.~Mehdizadeh, W.~Demtröder, Th. Zimmermann, H.~KÖppel, and L.~S.
  Cederbaum.
\newblock Vibronic {Level} {Density} of {Excited} {NO2}-{States} and its
  {Statistical} {Analysis}.
\newblock \emph{Berichte der Bunsengesellschaft für physikalische Chemie},
  92\penalty0 (3):\penalty0 312--318, 1988.
\newblock ISSN 0005-9021.
\newblock \doi{10.1002/bbpc.198800067}.
\newblock URL
  \url{https://onlinelibrary.wiley.com/doi/abs/10.1002/bbpc.198800067}.

\bibitem[Polik et~al.(1990)Polik, Guyer, Miller, and
  Moore]{polik_eigenstateresolved_1990}
William~F. Polik, Dean~R. Guyer, William~H. Miller, and C.~Bradley Moore.
\newblock Eigenstate‐resolved unimolecular reaction dynamics: {Ergodic}
  character of {S} formaldehyde at the dissociation threshold.
\newblock \emph{The Journal of Chemical Physics}, 92\penalty0 (6):\penalty0
  3471--3484, 1990.
\newblock ISSN 0021-9606.
\newblock \doi{10.1063/1.457858}.
\newblock URL \url{https://doi.org/10.1063/1.457858}.

\bibitem[Zimmermann et~al.(1988)Zimmermann, Köppel, Cederbaum, Persch, and
  Demtröder]{zimmermann_confirmation_1988}
Th. Zimmermann, H.~Köppel, L.~S. Cederbaum, G.~Persch, and W.~Demtröder.
\newblock Confirmation of random-matrix fluctuations in molecular spectra.
\newblock \emph{Phys. Rev. Lett.}, 61\penalty0 (1):\penalty0 3--6, 1988.
\newblock ISSN 0031-9007.
\newblock \doi{10.1103/PhysRevLett.61.3}.
\newblock URL \url{https://link.aps.org/doi/10.1103/PhysRevLett.61.3}.

\bibitem[Leitner et~al.(1996)Leitner, Köppel, and
  Cederbaum]{leitner_statistical_1996}
David~M. Leitner, H.~Köppel, and L.~S. Cederbaum.
\newblock Statistical properties of molecular spectra and molecular dynamics:
  {Analysis} of their correspondence in {NO2} and {C2H}+4.
\newblock \emph{The Journal of Chemical Physics}, 104\penalty0 (2):\penalty0
  434--443, 1996.
\newblock ISSN 0021-9606.
\newblock \doi{10.1063/1.470842}.
\newblock URL \url{https://doi.org/10.1063/1.470842}.

\bibitem[Takatsuka(2022)]{takatsuka_quantum_2022}
Kazuo Takatsuka.
\newblock Quantum chaos in the dynamics of molecules.
\newblock \emph{Entropy}, 25\penalty0 (1):\penalty0 63, 2022.
\newblock ISSN 1099-4300.
\newblock \doi{10.3390/e25010063}.
\newblock URL \url{https://www.mdpi.com/1099-4300/25/1/63}.

\bibitem[Robinson and Holbrook(1972)]{RobinsonHolbrook1972}
P.~J. Robinson and K.~A. Holbrook.
\newblock \emph{Unimolecular Reactions}.
\newblock Wiley-Interscience, New York, 1972.

\bibitem[Wolynes(1992)]{wolynes_randomness_1992}
Peter~G. Wolynes.
\newblock Randomness and complexity in chemical physics.
\newblock \emph{Accounts of Chemical Research}, 25\penalty0 (11):\penalty0
  513--519, 1992.
\newblock ISSN 0001-4842.
\newblock \doi{10.1021/ar00023a005}.
\newblock URL \url{https://doi.org/10.1021/ar00023a005}.

\bibitem[Leitner(2015)]{leitner_quantum_2015}
David~M. Leitner.
\newblock Quantum ergodicity and energy flow in molecules.
\newblock \emph{Advances in Physics}, 64\penalty0 (4):\penalty0 445--517, 2015.
\newblock ISSN 0001-8732.
\newblock URL \url{https://doi.org/10.1080/00018732.2015.1109817}.

\bibitem[Leitner(2018)]{leitner_molecules_2018}
David~M. Leitner.
\newblock Molecules and the {Eigenstate} {Thermalization} {Hypothesis}.
\newblock \emph{Entropy}, 20\penalty0 (9), 2018.
\newblock ISSN 1099-4300.
\newblock \doi{10.3390/e20090673}.
\newblock URL \url{https://www.mdpi.com/1099-4300/20/9/673}.

\bibitem[Born and Oppenheimer(1927)]{Born1927}
M.~Born and R.~Oppenheimer.
\newblock Zur {Quantentheorie} der {Molekeln}.
\newblock \emph{Annalen der Physik}, 389\penalty0 (20):\penalty0 457--484,
  1927.
\newblock \doi{10.1002/andp.19273892002}.

\bibitem[Born et~al.(1955)Born, Huang, and Lax]{Born1955}
Max Born, Kun Huang, and M~Lax.
\newblock Dynamical theory of crystal lattices.
\newblock \emph{American Journal of Physics}, 23\penalty0 (7):\penalty0
  474--474, 1955.
\newblock \doi{10.1119/1.1934059}.

\bibitem[Szabo and Ostlund(1989)]{szabo_modern_1989}
Attila Szabo and Neil~S. Ostlund.
\newblock \emph{Modern Quantum Chemistry}.
\newblock Dover Publications, Inc., 1989.
\newblock ISBN 978-0-486-13459-8.

\bibitem[Berry(1977)]{BerryC}
M.~V. Berry.
\newblock Regular and irregular semiclassical wavefunctions.
\newblock \emph{Journal of Physics A: Mathematical and General}, 10\penalty0
  (12):\penalty0 2083--2091, 1977.
\newblock \doi{10.1088/0305-4470/10/12/016}.

\bibitem[Bunimovich(1974)]{Bunimovich}
L.~A. Bunimovich.
\newblock On ergodic properties of certain billiards.
\newblock \emph{Functional Analysis and Its Applications}, 8\penalty0
  (3):\penalty0 254--255, 1974.
\newblock \doi{10.1007/BF01075700}.

\bibitem[Rozenbaum et~al.(2020)Rozenbaum, Bunimovich, and
  Galitski]{BunimovichGalitski}
Efim~B. Rozenbaum, Leonid~A. Bunimovich, and Victor Galitski.
\newblock Early-time exponential instabilities in nonchaotic quantum systems.
\newblock \emph{Physical Review Letters}, 125:\penalty0 014101, 2020.
\newblock \doi{10.1103/PhysRevLett.125.014101}.

\bibitem[Heller(1984)]{Scars}
Eric~J. Heller.
\newblock Bound-state eigenfunctions of classically chaotic hamiltonian
  systems: Scars of periodic orbits.
\newblock \emph{Physical Review Letters}, 53\penalty0 (16):\penalty0
  1515--1518, 1984.
\newblock \doi{10.1103/PhysRevLett.53.1515}.

\bibitem[Dreuw and Head-Gordon(2005)]{dreuw_single-reference_2005}
Andreas Dreuw and Martin Head-Gordon.
\newblock Single-reference ab initio methods for the calculation of excited
  states of large molecules.
\newblock \emph{Chem. Rev.}, 105\penalty0 (11):\penalty0 4009--4037, 2005.
\newblock ISSN 0009-2665.
\newblock \doi{10.1021/cr0505627}.
\newblock URL \url{https://doi.org/10.1021/cr0505627}.

\bibitem[Liao et~al.(2020)Liao, Vikram, and Galitski]{LiaoVikramGalitski}
Yunxiang Liao, Amit Vikram, and Victor Galitski.
\newblock Many-body level statistics of single-particle quantum chaos.
\newblock \emph{Physical Review Letters}, 125:\penalty0 250601, 2020.
\newblock \doi{10.1103/PhysRevLett.125.250601}.

\bibitem[Deutsch(1991)]{ETH1}
J.~M. Deutsch.
\newblock Quantum statistical mechanics in a closed system.
\newblock \emph{Phys. Rev. A}, 43:\penalty0 2046--2049, 1991.

\bibitem[Srednicki(1994)]{ETH2}
M.~Srednicki.
\newblock Chaos and quantum thermalization.
\newblock \emph{Phys. Rev. E}, 50:\penalty0 888--901, 1994.

\bibitem[D'Alessio et~al.(2016)D'Alessio, Kafri, Polkovnikov, and
  Rigol]{MarcosTolya}
L.~D'Alessio, Y.~Kafri, A.~Polkovnikov, and M.~Rigol.
\newblock From quantum chaos and eigenstate thermalization to statistical
  mechanics and thermodynamics.
\newblock \emph{Adv. Phys.}, 65:\penalty0 239--362, 2016.

\bibitem[Prange(1997)]{Prange}
R.~E. Prange.
\newblock The spectral form factor is not self-averaging.
\newblock \emph{Physical Review Letters}, 78:\penalty0 2280--2283, 1997.
\newblock \doi{10.1103/PhysRevLett.78.2280}.

\bibitem[Vikram and Galitski(2024)]{Bound}
Amit Vikram and Victor Galitski.
\newblock Exact universal bounds on quantum dynamics and fast scrambling.
\newblock \emph{Physical Review Letters}, 132:\penalty0 040402, 2024.
\newblock \doi{10.1103/PhysRevLett.132.040402}.

\bibitem[Reimann(2016)]{Reimann}
Peter Reimann.
\newblock Typical fast thermalization processes in closed many-body systems.
\newblock \emph{Nature Communications}, 7:\penalty0 10821, 2016.
\newblock \doi{10.1038/ncomms10821}.

\bibitem[Berry(1985)]{BerryRigidity}
M.~V. Berry.
\newblock Semiclassical theory of spectral rigidity.
\newblock \emph{Proceedings of the Royal Society of London A}, 400:\penalty0
  229--251, 1985.
\newblock \doi{10.1098/rspa.1985.0050}.

\bibitem[Liao and Galitski(2022)]{Dephasing}
Yunxiang Liao and Victor Galitski.
\newblock Universal dephasing mechanism of many-body quantum chaos.
\newblock \emph{Physical Review Research}, 4:\penalty0 L012037, 2022.
\newblock \doi{10.1103/PhysRevResearch.4.L012037}.

\bibitem[Vikram et~al.(2026)Vikram, Shou, and Galitski]{SpeedLimit}
Amit Vikram, Laura Shou, and Victor Galitski.
\newblock Proof of a universal speed limit on fast scrambling in quantum
  systems.
\newblock \emph{Physical Review Letters}, 2026.
\newblock To appear.

\bibitem[Argaman et~al.(1993)Argaman, Dittes, Doron, Keating, Kitaev, Sieber,
  and Smilansky]{POrbits}
N.~Argaman, F.-M. Dittes, E.~Doron, J.~P. Keating, A.~Y. Kitaev, M.~Sieber, and
  U.~Smilansky.
\newblock Correlations in the actions of periodic orbits derived from quantum
  chaos.
\newblock \emph{Physical Review Letters}, 71:\penalty0 4326--4329, 1993.
\newblock \doi{10.1103/PhysRevLett.71.4326}.

\bibitem[Altland and Kamenev(2000)]{AltlandKamenev}
Alexander Altland and Alex Kamenev.
\newblock Wigner-dyson statistics from the keldysh $\sigma$-model.
\newblock \emph{Physical Review Letters}, 85:\penalty0 5615--5618, 2000.
\newblock \doi{10.1103/PhysRevLett.85.5615}.

\bibitem[Montgomery(1973)]{Montgomery}
Hugh~L. Montgomery.
\newblock The pair correlation of zeros of the zeta function.
\newblock In \emph{Analytic Number Theory}, volume~24 of \emph{Proceedings of
  Symposia in Pure Mathematics}, pages 181--193. American Mathematical Society,
  Providence, RI, 1973.

\bibitem[London(1937)]{london_theorie_1937}
F.~London.
\newblock Théorie quantique des courants interatomiques dans les combinaisons
  aromatiques.
\newblock \emph{Journal de Physique et le Radium}, 8\penalty0 (10):\penalty0
  397--409, 1937.
\newblock \doi{10.1051/jphysrad:01937008010039700}.
\newblock URL \url{https://hal.science/jpa-00233534}.

\bibitem[Helgaker and Jo/rgensen(1991)]{helgaker_electronic_1991}
Trygve Helgaker and Poul Jo/rgensen.
\newblock An electronic hamiltonian for origin independent calculations of
  magnetic properties.
\newblock \emph{J. Chem. Phys.}, 95\penalty0 (4):\penalty0 2595--2601, 1991.
\newblock ISSN 0021-9606.
\newblock \doi{10.1063/1.460912}.
\newblock URL \url{https://doi.org/10.1063/1.460912}.

\bibitem[Simons et~al.(1993)Simons, Hashimoto, Courtney, Kleppner, and
  Altshuler]{MHydrogen}
B.~D. Simons, A.~Hashimoto, M.~Courtney, D.~Kleppner, and B.~L. Altshuler.
\newblock New class of universal correlations in the spectra of hydrogen in a
  magnetic field.
\newblock \emph{Physical Review Letters}, 71\penalty0 (18):\penalty0
  2899--2902, 1993.
\newblock \doi{10.1103/PhysRevLett.71.2899}.

\bibitem[Wang et~al.(2013)Wang, Steinberg, Jarillo-Herrero, and Gedik]{Gedik}
Y.~H. Wang, H.~Steinberg, P.~Jarillo-Herrero, and N.~Gedik.
\newblock Observation of floquet-bloch states on the surface of a topological
  insulator.
\newblock \emph{Science}, 342\penalty0 (6157):\penalty0 453--457, 2013.
\newblock \doi{10.1126/science.1239834}.

\bibitem[Lindner et~al.(2011)Lindner, Refael, and Galitski]{FTI}
Netanel~H. Lindner, Gil Refael, and Victor Galitski.
\newblock Floquet topological insulator in semiconductor quantum wells.
\newblock \emph{Nat. Phys.}, 7:\penalty0 490--495, 2011.
\newblock \doi{10.1038/nphys1926}.

\bibitem[Floquet(1883)]{Floquet}
Gaston Floquet.
\newblock Sur les {\'e}quations diff{\'e}rentielles lin{\'e}aires {\`a}
  coefficients p{\'e}riodiques.
\newblock \emph{Annales Scientifiques de l'{\'E}cole Normale Sup{\'e}rieure},
  12:\penalty0 47--88, 1883.
\newblock \doi{10.24033/asens.220}.

\bibitem[Zhou et~al.(2023)Zhou, Wu, Bian, and Subotnik]{SubotnikFloquet}
Z.~Zhou, Y.~Wu, X.~Bian, and J.~E. Subotnik.
\newblock Nonadiabatic dynamics in a continuous circularly polarized laser
  field with floquet phase-space surface hopping.
\newblock \emph{J. Chem. Theory Comput.}, 19\penalty0 (3):\penalty0 718--732,
  2023.
\newblock ISSN 1549-9618.
\newblock \doi{10.1021/acs.jctc.2c00948}.

\bibitem[Simons and Altshuler(1993)]{SimonsAltshuler}
B.~D. Simons and B.~L. Altshuler.
\newblock Universalities in the spectra of disordered and chaotic systems.
\newblock \emph{Physical Review B}, 48\penalty0 (8):\penalty0 5422--5431, 1993.
\newblock \doi{10.1103/PhysRevB.48.5422}.

\bibitem[Zakrzewski and Delande(1993)]{ZD1}
Jakub Zakrzewski and Dominique Delande.
\newblock Parametric motion of energy levels in quantum chaotic systems. {I}.
  curvature distributions.
\newblock \emph{Physical Review E}, 47:\penalty0 1650--1663, 1993.
\newblock \doi{10.1103/PhysRevE.47.1650}.

\bibitem[Zakrzewski et~al.(1993)Zakrzewski, Delande, and Ku{\'s}]{ZD2}
Jakub Zakrzewski, Dominique Delande, and Marek Ku{\'s}.
\newblock Parametric motion of energy levels in quantum chaotic systems. {II}.
  avoided-crossing distributions.
\newblock \emph{Physical Review E}, 47:\penalty0 1665--1675, 1993.
\newblock \doi{10.1103/PhysRevE.47.1665}.

\bibitem[von Oppen(1994)]{vonOppen}
Felix von Oppen.
\newblock Exact distribution of eigenvalue curvatures of chaotic quantum
  systems.
\newblock \emph{Physical Review Letters}, 73:\penalty0 798--801, 1994.
\newblock \doi{10.1103/PhysRevLett.73.798}.

\bibitem[Fyodorov and Sommers(1995)]{FyodorovSommers}
Yan~V. Fyodorov and Hans-J{\"u}rgen Sommers.
\newblock ``level curvature'' distribution for diffusive {A}haronov--{B}ohm
  systems: Analytical results.
\newblock \emph{Physical Review E}, 51:\penalty0 R2719--R2722, 1995.
\newblock \doi{10.1103/PhysRevE.51.R2719}.

\bibitem[Domcke et~al.(2004)Domcke, Yarkony, and Köppel]{Con1}
Wolfgang Domcke, David~R. Yarkony, and Horst Köppel.
\newblock \emph{Conical Intersections}.
\newblock World Scientific, 2004.
\newblock \doi{10.1142/5406}.

\bibitem[Domcke and Yarkony(2012)]{Con2}
Wolfgang Domcke and David~R. Yarkony.
\newblock Role of conical intersections in molecular spectroscopy and
  photoinduced chemical dynamics.
\newblock \emph{Annual Review of Physical Chemistry}, 63:\penalty0 325--352,
  2012.
\newblock \doi{10.1146/annurev-physchem-032511-143631}.

\bibitem[Berry and Wilkinson(1984)]{Diablo}
M.~V. Berry and M.~Wilkinson.
\newblock Diabolical points in the spectra of triangles.
\newblock \emph{Proceedings of the Royal Society of London A}, 392:\penalty0
  15--43, 1984.
\newblock \doi{10.1098/rspa.1984.0022}.

\bibitem[Kitaev(2009)]{KitaevTI}
Alexei Kitaev.
\newblock Periodic table for topological insulators and superconductors.
\newblock In \emph{AIP Conference Proceedings}, volume 1134, pages 22--30.
  American Institute of Physics, 2009.
\newblock \doi{10.1063/1.3149495}.

\bibitem[Hirata and Head-Gordon(1999)]{hirata_time-dependent_1999}
So~Hirata and Martin Head-Gordon.
\newblock Time-dependent density functional theory within the tamm–dancoff
  approximation.
\newblock \emph{Chemical Physics Letters}, 314\penalty0 (3):\penalty0 291--299,
  1999.
\newblock ISSN 0009-2614.
\newblock \doi{10.1016/S0009-2614(99)01149-5}.
\newblock URL
  \url{https://www.sciencedirect.com/science/article/pii/S0009261499011495}.

\bibitem[Epifanovsky et~al.(2021)Epifanovsky, Gilbert, Feng, Lee, Mao,
  et~al.]{Epifanovsky2021}
Evgeny Epifanovsky, Andrew T.~B. Gilbert, Xintian Feng, Joonho Lee, Yuezhi Mao,
  et~al.
\newblock Software for the frontiers of quantum chemistry: An overview of
  developments in the q-chem 5 package.
\newblock \emph{The Journal of Chemical Physics}, 155\penalty0 (8):\penalty0
  084801, 2021.
\newblock \doi{10.1063/5.0055522}.

\bibitem[D'Alessandro and Genovese(2019)]{dalessandro_locality_2019}
Marco D'Alessandro and Luigi Genovese.
\newblock Locality and computational reliability of linear response
  calculations for molecular systems.
\newblock \emph{Phys. Rev. Mater.}, 3\penalty0 (2):\penalty0 023805, 2019.

\bibitem[Otero et~al.(2019)Otero, Karamanis, and Mandado]{otero_new_2019}
Nicolás Otero, Panaghiotis Karamanis, and Marcos Mandado.
\newblock A new method to analyze and understand molecular linear and nonlinear
  optical responses via field-induced functions: a straightforward alternative
  to sum-over-states ({SOS}) analysis.
\newblock \emph{Physical Chemistry Chemical Physics}, 21\penalty0
  (11):\penalty0 6274--6286, 2019.
\newblock ISSN 1463-9084.
\newblock \doi{10.1039/C8CP07362G}.
\newblock URL
  \url{https://pubs.rsc.org/en/content/articlelanding/2019/cp/c8cp07362g}.

\bibitem[Hohm(2013)]{hohm_experimental_2013}
U.~Hohm.
\newblock Experimental static dipole–dipole polarizabilities of molecules.
\newblock \emph{Journal of Molecular Structure}, 1054-1055:\penalty0 282--292,
  December 2013.
\newblock ISSN 0022-2860.
\newblock \doi{10.1016/j.molstruc.2013.10.003}.
\newblock URL
  \url{https://www.sciencedirect.com/science/article/pii/S002228601300834X}.

\bibitem[Tracy and Widom(1994)]{TW1}
Craig~A. Tracy and Harold Widom.
\newblock Level-spacing distributions and the airy kernel.
\newblock \emph{Communications in Mathematical Physics}, 159:\penalty0
  151--174, 1994.
\newblock \doi{10.1007/BF02100489}.

\bibitem[Tracy and Widom(1996)]{TW2}
Craig~A. Tracy and Harold Widom.
\newblock On orthogonal and symplectic matrix ensembles.
\newblock \emph{Communications in Mathematical Physics}, 177:\penalty0
  727--754, 1996.
\newblock \doi{10.1007/BF02099545}.

\bibitem[González et~al.(2012)González, Escudero, and
  Serrano-Andrés]{gonzalez_progress_2012}
Leticia González, Daniel Escudero, and Luis Serrano-Andrés.
\newblock Progress and {Challenges} in the {Calculation} of {Electronic}
  {Excited} {States}.
\newblock \emph{ChemPhysChem}, 13\penalty0 (1):\penalty0 28--51, 2012.
\newblock ISSN 1439-7641.
\newblock \doi{10.1002/cphc.201100200}.
\newblock URL
  \url{https://onlinelibrary.wiley.com/doi/abs/10.1002/cphc.201100200}.
\newblock \_eprint:
  https://chemistry-europe.onlinelibrary.wiley.com/doi/pdf/10.1002/cphc.201100200.

\bibitem[Lischka et~al.(2018)Lischka, Nachtigallová, Aquino, Szalay, Plasser,
  Machado, and Barbatti]{lischka_multireference_2018}
Hans Lischka, Dana Nachtigallová, Adélia J.~A. Aquino, Péter~G. Szalay,
  Felix Plasser, Francisco B.~C. Machado, and Mario Barbatti.
\newblock Multireference {Approaches} for {Excited} {States} of {Molecules}.
\newblock \emph{Chemical Reviews}, 118\penalty0 (15):\penalty0 7293--7361,
  2018.
\newblock ISSN 0009-2665.
\newblock \doi{10.1021/acs.chemrev.8b00244}.
\newblock URL \url{https://doi.org/10.1021/acs.chemrev.8b00244}.

\bibitem[Abedi et~al.(2012)Abedi, Maitra, and Gross]{Abedi2012}
Ali Abedi, Neepa~T. Maitra, and E.~K.~U. Gross.
\newblock Correlated electron-nuclear dynamics: Exact factorization of the
  molecular wavefunction.
\newblock \emph{The Journal of Chemical Physics}, 137\penalty0 (22):\penalty0
  22A530, 2012.
\newblock \doi{10.1063/1.4745836}.

\bibitem[Agostini and Curchod(2022)]{agostini_chemistry_2022}
Federica Agostini and Basile F.~E. Curchod.
\newblock Chemistry without the {Born}–{Oppenheimer} approximation.
\newblock \emph{Philosophical Transactions of the Royal Society A:
  Mathematical, Physical and Engineering Sciences}, 380\penalty0
  (2223):\penalty0 20200375, 2022.
\newblock ISSN 1364-503X.
\newblock \doi{10.1098/rsta.2020.0375}.
\newblock URL \url{https://doi.org/10.1098/rsta.2020.0375}.

\bibitem[Bian et~al.(2026)Bian, Duston, Bradbury, Tao, Bhati, Qiu, Wu, Wu, and
  Subotnik]{bian_phase-space_2026}
Xuezhi Bian, Titouan Duston, Nadine Bradbury, Zhen Tao, Mansi Bhati, Tian Qiu,
  Xinchun Wu, Yanze Wu, and Joseph~E. Subotnik.
\newblock The phase-space way to electronic structure theory and subsequently
  chemical dynamics.
\newblock \emph{Chemical Physics Reviews}, 7\penalty0 (1):\penalty0 011303,
  2026.
\newblock ISSN 2688-4070.
\newblock \doi{10.1063/5.0286240}.
\newblock URL \url{https://doi.org/10.1063/5.0286240}.

\end{thebibliography}

\end{document}